\documentclass[english,eqsecnum,twocolumn]{revtex4}
\usepackage{mathptmx}
\usepackage{helvet}
\usepackage{courier}
\usepackage[T1]{fontenc}
\usepackage[latin1]{inputenc}
\usepackage{bm}
\usepackage{graphicx}

\makeatletter

\usepackage{bm}

\makeatletter

\makeatletter

\usepackage{mathptmx}

\usepackage{helvet}

\usepackage{courier}



\newcommand{\Tr}{\text{Tr}}

\makeatother

\makeatother

\makeatother

\usepackage{babel}

\begin{document}

\title{Qubit phase-space: SU($n$) coherent state P-representations}

\author{D. W. Barry and P. D. Drummond}

\affiliation{ARC Centre of Excellence for Quantum-Atom Optics, School of Physical
Sciences, University of Queensland, Brisbane, Queensland 4072, Australia}

\begin{abstract}
We introduce a phase-space representation for qubits and spin models.
The technique uses an SU(n) coherent state basis, and can equally
be used for either static or dynamical simulations. We review previously
known definitions and operator identities, and show how these can
be used to define an off-diagonal, positive phase-space representation
analogous to the positive P-function. As an illustration of the phase-space
method, we use the example of the Ising model, which has exact solutions
for the finite temperature canonical ensemble in two dimensions. We
show how a canonical ensemble for an Ising model of arbitrary structure
can be efficiently simulated using SU(2) or atomic coherent states.
The technique utilizes a transformation from a canonical (imaginary-time)
weighted simulation to an equivalent unweighted real-time simulation.
The results are compared to the exactly soluble two-dimensional case.
We note that Ising models in one, two or three dimensions are potentially
achievable experimentally as a lattice-gas of ultra-cold atoms in
optical lattices. The technique is not restricted to canonical ensembles
or to Ising-like couplings. It is also able to be used for real-time
evolution, and for systems whose time-evolution follows a master-equation
describing decoherence and coupling to external reservoirs. The case
of SU(n) phase-space is used to describe $n$-level systems. In general,
the requirement that time-evolution is stochastic corresponds to a
restriction to Hamiltonians and master-equations that are quadratic
in the group generators or generalized spin operators. 
\end{abstract}
\maketitle

\section{Introduction}

Qubits are a central concept in quantum information. However, complexity
issues mean that calculations with large numbers of qubits are nontrivial:
the Hilbert space dimension scales as $2^{M}$ for $M$ qubits. A
natural way to treat this type of complexity is to use a phase-space
representation over an atomic coherent state basis. Coherent states,
introduced by Schrodinger\cite{Schrod}, have been used widely in
quantum optics. Atomic coherent states -- originally used for collections
of two-level atoms\cite{Dicke} -- are the natural solution for a
quantum spin driven by an external driving force, like a magnetic
field. They are also called SU(2) \cite{Radcliffe,ACGT}, spin, or
more generally SU($n$) coherent states\cite{Perelemov,Gilmore}
for arbitrary $n$-level systems. Since they are a continuous set,
they satisfy differential identities, which can have useful applications.

In this paper, a phase-space representation of arbitrary density matrices
in terms of off-diagonal SU($n$) coherent state projectors is introduced.
This extends earlier P-function\cite{Narducci} and Q-function\cite{Gilmore,Lee}
approaches involving SU(2) and SU($n$) projectors\cite{Drummond_SU(N)}.
The methods described here allow dynamical or static entanglement
to be treated, and extend earlier phase-space approaches in quantum
optics\cite{Wig-Wigner,Hus-Q,GSp,CG-Q,Agarwal}. In particular, they
include off-diagonal coherent-state projectors which lead to positive-definite
diffusion, and hence to dynamical realisations as stochastic processes\cite{positiveP1,pp,QuantNoise}.
The resulting methods have applications to either time-evolution or
canonical ensemble calculations of finite Hilbert space systems with
spin systems. More general applications in quantum information are
also possible, owing to the simplicity with which large and/or decoherent
spin systems can be treated.

Other methods for treating finite Hilbert spaces like coupled spins
include finite versions of the Wigner representation\cite{ChatWig},
path-integral techniques\cite{Pathint} and DMRG-based methods \cite{white92,vidal03,verstraete04}.
While these are interesting and often very useful, they are not suited
to exact, probabilistic simulations, because they either involve approximations,
or else they do not use a positive distribution function. When DMRG
techniques are possible - typically in one-dimensional ground-state
calculations - they are very accurate and useful, but this method
often cannot be used in many other physical examples involving finite
temperatures, dissipation, dynamics or higher dimensions.

Exact methods also exist - like the one and two dimensional Ising
model at finite temperature - but these approaches are restricted
to special cases. Our approach is to define a positive distribution
function over a space of SU($n$) coherent state amplitudes. This
is a much smaller dimension than the whole Hilbert space, scaling
proportionally to the number of spins. We emphasize that the representation
is not unique, and some care is needed in choosing the expansion to
minimise sampling error. In general, the main restriction is the compactness
or otherwise of the resulting phase-space distribution: if there are
large distribution variances, this will increase sampling error in
a practical calculation.

As an example to illustrate scaling behaviour in an exactly soluble
case, the application of SU(2) or atomic coherent states to solving
the two-dimensional Ising model is treated in detail. This application
is simple yet instructive, and the resulting algorithm is novel and
efficient. The Ising model\cite{Ising} is one of the oldest models
in statistical mechanics, with many applications\cite{Landau}. The
model has the virtue of having a non-trivial exact solution in two-dimensions\cite{Onsager,Baxter}.
It displays a critical-point phase-transition\cite{Fisher}, which
we use to test the phase-space method. We find excellent agreement
with these exact results.

The original use of the Ising model was a simple theory of ferromagnetism
-- in which atomic spins have either an `up' or `down' orientation.
It also finds applications to a variety of other physical problems,
from the theory of lattice gases and binary alloys to spin glasses\cite{BinderYoung},
percolation\cite{Percolation} and other disordered systems. Modern
ultra-cold atom experiments with optical lattices\cite{Jaksch} can
test this model directly, at temperatures above quantum degeneracy
where the lattice-gas model is applicable. In this case, the two states
of each lattice site correspond simply to the presence or absence
of a single atom. At lower temperatures where coherences are important,
Heisenberg-like models become applicable, and these will be treated
elsewhere.

There are numerous corresponding techniques for solving the Ising
model. However, exact solutions are known only in special cases like
the uniform one and two-dimensional lattices. More generally, the
other techniques that are known rely on Monte-Carlo methods\cite{Binder,Swendsen,BloteHeringa},
in which the space of all configurations is searched by random spin-flipping
algorithms\cite{Metropolis}. The method demonstrated here is quite
different to traditional approaches.

The SU($n$) phase-space approach can also be readily used for other
models of interacting spins, to real time evolution and to dynamical
couplings to reservoirs, where no exact solution is known. While these
applications will be treated elsewhere, we note that the main restrictions
are that the Hamiltonian or master equation should be at most quadratic
in the SU($n$) operators, which is the typical case for coupled spin
systems. It is intriguing to note that these types of problems are
also regarded as potentially soluble for future generations of quantum
computers. The methods proposed here have the advantage that they
can be implemented on digital computers. Thus, they complement the
quantum computing approach, and indeed can be used to simulate quantum
logic gates in the presence of decoherence. The main limiting is sampling
error, which typically grows with simulation time.

\section{SU(2) coherent states}

We start with the well-known SU(2) case, which corresponds to a spin-J
physical system or more generally, a collection of physically equivalent
two-level systems. The SU(2) coherent states or atomic coherent states
are defined for states generated with angular momentum raising and
lowering operators\cite{Radcliffe,ACGT}. These are physically important
in many systems, ranging from groups of two-level atoms to nuclear
spins, as well as superconducting qubits and other systems with an
SU(2) symmetry.

The relevant spin operators $\hat{S}$ have commutators defined so
that:

\begin{equation}
\left[\hat{S}_{i},\hat{S}_{j}\right]=i\epsilon_{ijk}\hat{S}_{k}.\end{equation}
 Here $\epsilon_{ijk}=\pm1$ depending on whether the indices are
in cyclic or anti-cyclic order, and one conventionally writes $\hat{S}_{x,y,z}$
to denote $\hat{S}_{1,2,3}$. It is useful to also define the raising
and lowering operators which act on an eigenstate of $\hat{S}_{z}$
to increase (decrease) the eigenvalue. These are defined as: \begin{equation}
\hat{S}^{\pm}=\hat{S}_{x}\pm i\hat{S}_{y}.\end{equation}

We consider a subsystem with a definite value of \begin{equation}
\hat{S}_{x}^{2}+\hat{S}_{y}^{2}+\hat{S}_{z}^{2}=\hat{S}^{2}=S(S+1)\,.\end{equation}
 Physically, these may be obtained either directly as an atom or molecule
of spin $S$, or equivalently from a grouping of $N\geq2S$ spin $1/2$
quantum systems or qubits, each with $2$-levels and equivalent couplings.
These composite systems in general have $2S+1$ distinct energy levels,
and there is a unique lowest eigenstate of $\hat{S}_{z}$, denoted
$\left|0\right\rangle $.

The standard definition of SU(2) coherent states\cite{Radcliffe,ACGT}
is that they are the states generated from $\left|0\right\rangle $
by the raising operator, so that, for a spin-$S$ basis,

\begin{equation}
\left|\alpha\right\rangle ^{(2)}=\frac{e^{\hat{S^{+}}\alpha}}{\left[1+\left|\alpha\right|^{2}\right]^{J}}\left|0\right\rangle \,.\label{eq:normACS}\end{equation}

It is convenient here to also consider an un-normalized version of
this atomic coherent state, which we define as

\begin{equation}
\left\Vert \overrightarrow{\psi}\right\rangle ^{(2)}=\left[\psi^{0}\right]^{N}e^{\psi^{1}\hat{S}^{+}/\psi^{0}}\left|0\right\rangle \label{eq:unnorm-op-ACS}\end{equation}
 For simplicity in obtaining identities, it is useful to have just
one complex parameter, as in the standard definition. Our choice is
to define \begin{eqnarray}
\psi^{1} & = & \exp(z/2)\nonumber \\
\psi^{0} & = & \exp(-z/2)\,,\end{eqnarray}
 where $z=r+i\phi=\ln\alpha$ is a complex parameter. With this choice,
the SU(2) coherent states are parametrized over a one-dimensional
complex manifold, or a two-dimensional real manifold. We will represent
this parametrization as $\left\Vert z\right\rangle $, where

\begin{eqnarray}
\left\Vert z\right\rangle  & = & e^{-Sz}e^{\hat{S}^{+}e^{z}}\left|0\right\rangle \,\,.\end{eqnarray}

For visualization purposes, one may project the atomic coherent state
phase-space onto a spherical surface, called the Bloch sphere. In
this case, it is usual to normalize the state, and to define \begin{equation}
\left|\theta,\phi\right\rangle ^{(2)}=\left|e^{i\phi}\tan\theta/2\right\rangle ^{(2)}\,\,.\label{eq:BlochACS}\end{equation}
 This Bloch-sphere mapping therefore involves the transformation of
\begin{equation}
\alpha=e^{i\phi}\tan\theta/2=e^{z}\,\,.\end{equation}

\subsubsection{Two-level case}

As an illustration of the simplest case possible, where $S=1/2,$
we consider a two-level Hilbert space having quantum states labelled
$\left|0\right\rangle $ and $\left|1\right\rangle $. This corresponds
to a single qubit in quantum information terminology. An atomic coherent
state or SU(2) coherent state is then just an arbitrary pure qubit
state:\begin{eqnarray}
\left\Vert \overrightarrow{\psi}\right\rangle ^{(2)} & = & \psi^{0}\left|0\right\rangle +\psi^{1}\left|1\right\rangle \nonumber \\
 & = & e^{-z/2}\left|0\right\rangle +e^{z/2}\left|1\right\rangle \,\,.\end{eqnarray}

This shows the utility of this parametrization: it displays a symmetry
between up and down states, which simply corresponds to changing the
sign of $z$. In a useful vector representation, one can write this
in an explicit form as\begin{equation}
\left\Vert \overrightarrow{\psi}\right\rangle ^{(2)}=\left[\begin{array}{c}
\psi^{1}\\
\psi^{0}\end{array}\right]\,\,.\end{equation}
 In this notation, the state $\left|1\right\rangle $ corresponds
to spin projection $m=1/2$. Similarly, the second entry or state
$\left|0\right\rangle $ corresponds to spin projection $m=-1/2$.
On the Bloch sphere, this corresponds to \begin{eqnarray}
\left|\theta,\phi\right\rangle ^{(2)} & = & \frac{\psi^{0}\left|0\right\rangle +\psi^{1}\left|1\right\rangle }{\sqrt{\left|\psi^{0}\right|^{2}+\left|\psi^{1}\right|^{2}}}\,.\label{eq:BlochACS2}\\
 & = & \cos\frac{\theta}{2}e^{-i\phi/2}\left|0\right\rangle +\sin\frac{\theta}{2}e^{i\phi/2}\left|1\right\rangle \,\end{eqnarray}

\subsection{Lattice atomic coherent states}

For $M$ distinct spins, particles, or lattice sites, where one may
wish to address or couple to them individually, one must have $M$
distinct spin operators. As noted above, each of these can describe
$N$ physical qubits.

The corresponding outer-product SU(2) coherent state is then:

\begin{equation}
\left|\underline{\alpha}\right\rangle ^{(2,M)}=\prod_{m=1}^{M}\left[\frac{e^{\hat{S_{m}^{+}}\alpha_{m}}}{\left[1+\left|\alpha_{m}\right|^{2}\right]^{S}}\right]\left|0\right\rangle \,.\label{eq:lattice-normACS}\end{equation}
 For $N=1$, the two-level or qubit case, we note that with $\mathbf{z}=\underline{z}=(z_{1}\ldots z_{M})$,
our un-normalized definition becomes;

\begin{eqnarray}
\left\Vert \mathbf{z}\right\rangle  & = & \left\Vert \underline{\overrightarrow{\psi}}\right\rangle ^{(2,M)}\label{eq:un-normACS_z}\\
 & = & \otimes_{m=1}^{M}\left[e^{-z_{m}/2}\left|0\right\rangle _{m}+e^{z_{m}/2}\left|1\right\rangle _{m}\right].\end{eqnarray}

In this notation, the inner product is\begin{equation}
\left\langle \mathbf{z}\right\Vert \left\Vert \mathbf{z}'\right\rangle =2^{M}\prod_{m=1}^{M}\cosh\left(\left[z'_{m}+z_{m}^{*}\right]/2\right)\,\,,\label{eq:innerprod2}\end{equation}
 and we can therefore introduce a normalized state denoted $\left|\mathbf{z}\right\rangle $,
where

\begin{equation}
\left|\mathbf{z}\right\rangle =\prod_{m=1}^{M}\frac{1}{\sqrt{2\cosh\left(r_{m}\right)}}\left\Vert \mathbf{z}\right\rangle \,\,.\label{eq:normACS2}\end{equation}

\section{SU($n$) coherent states}

In cases where SU(2) symmetry does not hold, the SU(2) coherent states
can be generalized to SU(n) coherent states which are generated using
operators with an SU($n$) operator algebra.

The SU($n$) group is the group of $n\times n$ unitary matrices with
unit determinant, and so provides the most general way to treat the
transformations of an $n$-level quantum system. Therefore, SU($n$)
coherent states provide a useful basis set for general multi-level
quantum systems like atoms or spins. In the following section we review
results for the SU(n) coherent states. We also consider the important
case of outer products of SU(n) coherent states, which are needed
for treating lattices.

In the simplest case $n$ corresponds to the number of distinct quantum
states or levels involved. More generally, $n$ simply labels a symmetry
group which can have a larger dimensional representation, just as
in the SU($2$) case.

These states are useful in treating, for example, an assembly of $n$
coupled Bose-Einstein condensates, $n$-level atoms, or photon states
with $0,1\ldots n-1$ photons per mode. The SU($n$) algebra is generated
by the $n^{2}-1$ independent operators which satisfy the commutation
relations\cite{Hamermesh}

\begin{equation}
\left[\hat{R}^{\mu\nu},\hat{R}^{\mu'\nu'}\right]=\delta^{\nu\mu'}\hat{R}^{\mu\nu'}-\delta^{\nu'\mu}\hat{R}^{\mu'\nu}\,,\end{equation}
 together with the constraint that $\sum\hat{R}^{\mu\mu}=\hat{1}$.
The SU(n) coherent states can also be written in the following convenient
form, using an un-normalized notation in analogy to Eq. (\ref{eq:unnorm-op-ACS}),
as:\begin{equation}
\left\Vert \overrightarrow{\psi}\right\rangle ^{(n)}=\left[\psi^{0}\right]^{N}e^{\sum_{\mu>0}\psi^{\mu}\hat{R}^{\mu0}/\psi^{0}}\left|0\right\rangle \label{eq:unnorm-op-SU(n)}\end{equation}

We can use a collection of N equivalent $n$-level quantum systems
with states $\left|\mu\right\rangle _{j}$ for $\mu=0,\ldots,n-1$,
and $j=1,\ldots N$ , to indicate the essential features of this approach.
In this case the SU(n) operator algebra representation is provided
by:

\begin{equation}
\hat{R}^{\mu\nu}=\sum_{j=1}^{N}\left|\mu\right\rangle _{j}\left\langle \nu\right|_{j}\,\,.\end{equation}

For this case of N equivalent $n$-level atomic or spin states, one
can then define an SU(n) coherent state directly in terms of the original
Bloch basis $\left|k\right\rangle _{j}$, as: \begin{equation}
\left\Vert \overrightarrow{\psi}\right\rangle ^{(n)}=\prod_{j=1}^{N}\left[\sum_{\mu=0}^{n-1}\psi^{\mu}\left|\mu\right\rangle _{j}\right]\label{eq:un-normSU(n)}\end{equation}
 The corresponding normalized state is then:\begin{equation}
\left|\overrightarrow{\psi}\right\rangle ^{(n)}=\frac{1}{\left|\overrightarrow{\psi}\right|^{N}}\prod_{j=1}^{N}\left[\sum_{\mu=0}^{n-1}\psi^{\mu}\left|\mu\right\rangle _{j}\right]\label{eq:norm-SU(n)}\end{equation}

In the normalized case it is common to take the first coefficient
to be unity, so that $\psi^{0}=1$, although other choices are possible.
In general there are $n-1$ independent complex amplitudes of physical
significance, since the overall phase and amplitude of a wave-function
is physically irrelevant.

\subsection{Lattice SU($n$) coherent states}

Lattice coherent states we introduced in a pioneering work of Shastri
et al\cite{Shastry}, to study the Heisenberg model of interacting
spins. In our notation, for SU(n) coherent states defined at multiple
sites on a lattice labelled $m$, we introduce:

\begin{equation}
\left\Vert \underline{\overrightarrow{\psi}}\right\rangle ^{(n,M)}=\otimes_{m=1}^{M}\left\Vert \overrightarrow{\psi_{m}}\right\rangle _{m}^{(n)}\,,\label{eq:un-latticenormSU(n)}\end{equation}
 or, in a matrix notation analogous to the two-level case - except
with n levels per site -

\begin{equation}
\left\Vert \underline{\overrightarrow{\psi}}\right\rangle ^{(n,M)}=\otimes_{m=1}^{M}\left[\begin{array}{c}
\psi_{m}^{n-1}\\
\psi_{m}^{n-2}\\
\vdots\\
\psi_{m}^{0}\end{array}\right]\,.\end{equation}
 These multiple SU($n$) coherent states have the following inner
products:\begin{equation}
\left\langle \underline{\overrightarrow{\psi}}\right\Vert \left\Vert \underline{\overrightarrow{\psi}}'\right\rangle ^{(n,M)}=\prod_{m=1}^{M}\left[\overrightarrow{\psi_{m}^{*}}\cdot\overrightarrow{\psi}'_{m}\right]^{N}\,\,.\label{eq:innerprodSU(n)}\end{equation}
 One can also introduce \textbf{normalized} SU(n) lattice coherent
states, where the normalization uses the distance measure \begin{equation}
\left|\overrightarrow{\psi_{m}}\right|=\sqrt{\overrightarrow{\psi_{m}^{*}}\cdot\overrightarrow{\psi_{m}}}\,.\end{equation}
 Hence:\begin{equation}
\left|\underline{\mathbf{\overrightarrow{\psi}}}\right\rangle ^{(n,M)}=\prod_{m=1}^{M}\frac{1}{\left|\overrightarrow{\psi_{m}}\right|^{N}}\left\Vert \overrightarrow{\psi_{m}}\right\rangle ^{(n)}\,\,.\label{eq:normSU(n)}\end{equation}

These kinds of states can be thought of as generalizations of the
harmonic-oscillator coherent states, in the sense that with the usual
harmonic-oscillator coherent states there are prescribed relationships
between the coefficients. In the SU($n$) case there is no fixed relationship
between coefficients, but there is a fixed upper bound to the quantum
number.

\section{Completeness and Identities}

\subsection{Completeness}

The spin coherent states form an over-complete basis. In the SU(2)
case with spin-S, the resolution of the identity is well-known\cite{ACGT},
and is given by

\begin{equation}
\hat{1}=\left(2J+1\right)\int\frac{d\Omega}{4\pi}\left|\theta,\phi\right\rangle \left\langle \theta,\phi\right|\,,\label{eq:Ident}\end{equation}
 where $d\Omega=d\cos\theta d\phi$ is the usual integration measure
for the solid angle in spherical coordinates. In the spin-half case
with $N=1$, this can be simplified further, as one obtains from the
$z$-parameter mapping that\begin{eqnarray}
\hat{1} & = & \int_{0}^{2\pi}\frac{d\theta}{2\pi}\left[e^{i\theta/2}\left|1\right\rangle +e^{-i\theta/2}\left|0\right\rangle \right]\left[e^{-i\theta/2}\left\langle 1\right|+e^{i\theta/2}\left\langle 0\right|\right]\nonumber \\
 & = & \int_{0}^{2\pi}\frac{d\theta}{2\pi}\left\Vert i\theta\right\rangle \left\langle i\theta\right\Vert \nonumber \\
 & = & \int_{0}^{2\pi}\frac{d\theta}{\pi}\left|i\theta\right\rangle \left\langle i\theta\right|\,\,.\label{eq:2-level-completeness}\end{eqnarray}

In the more general SU($n$) case, one finds that\cite{Drummond_SU(N),GitmanSU(n)Completeness}:

\begin{equation}
\hat{1}=\frac{(N+n-1)!}{N!(2\pi)^{n}}\int\delta(\left|\overrightarrow{\psi}\right|^{2}-1)\left|\overrightarrow{\psi}\right\rangle \left\langle \overrightarrow{\psi}\right|d^{2n}\overrightarrow{\psi}\,.\label{eq:Ident_SU(n)}\end{equation}
 An even simpler resolution of the identity operator (for $N=1$)
is easily obtained with a multiple phase integration:\begin{eqnarray}
\hat{1} & = & \left|1\right\rangle \left\langle 1\right|+\left|2\right\rangle \left\langle 2\right|+\ldots+\left|n\right\rangle \left\langle n\right|\nonumber \\
 & = & \int_{0}^{2\pi}\ldots\int_{0}^{2\pi}\frac{d^{n-1}\underline{\theta}}{(2\pi)^{n-1}}\left[\sum e^{i\theta_{\mu}}\left|\mu\right\rangle \right]\left[\sum e^{-i\theta_{\mu}}\left\langle \mu\right|\right]\nonumber \\
 & = & \int_{0}^{2\pi}\ldots\int_{0}^{2\pi}\frac{d^{n-1}\underline{\theta}}{(2\pi)^{n-1}}\left\Vert \underline{e^{i\theta}}\right\rangle ^{(n)}\left\langle \underline{e^{i\theta}}\right\Vert ^{(n)}\,\,.\label{eq:completeness}\end{eqnarray}
 Just as in the two-level case, the first phase integral is omitted
here (ie, $\theta_{0}=0$), since this term is always orthogonal to
the others, due to the remaining phase-integrals.

\subsection{SU(n) operator identities}

We wish to obtain differential identities that involve the set of
operators that can act on the spin coherent states. These can all
be regarded as extensions of the very simple differential identities
that exist for the SU(n) coherent states. From Eq (\ref{eq:un-normSU(n)}),
one can directly prove that:\begin{equation}
\hat{R_{m}^{\mu\nu}}\left\Vert \underline{\overrightarrow{\psi}}\right\rangle ^{(n,M)}=\psi_{m}^{\nu}\frac{\partial}{\partial\psi_{m}^{\mu}}\left\Vert \underline{\overrightarrow{\psi}}\right\rangle ^{(n,M)}\,.\end{equation}

We now specialize to the two-level case where `raising' and `lowering'
operators are conventionally defined in physics as the matrices:\begin{equation}
\widehat{\sigma}^{+}=\left[\begin{array}{cc}
0 & 1\\
0 & 0\end{array}\right];\,\,\,\,\widehat{\sigma}^{-}=\left[\begin{array}{cc}
0 & 0\\
1 & 0\end{array}\right].\end{equation}
 These have a direct relationship with the $\hat{R}$ operators, since
for SU(2) symmetry with $S=1/2,$ one has: $\hat{R}^{01}=\hat{\sigma^{-}}$
and $\hat{R}^{10}=\hat{\sigma^{+}}$. In addition, $\widehat{\sigma}^{x,y,z}$
are the Pauli spin operators defined as:\begin{equation}
\widehat{\sigma}^{x}=\left[\begin{array}{cc}
0 & 1\\
1 & 0\end{array}\right],\,\,\,\,\widehat{\sigma}^{y}=\left[\begin{array}{cc}
0 & -i\\
i & 0\end{array}\right],\,\,\,\widehat{\sigma}^{z}=\left[\begin{array}{cc}
1 & 0\\
0 & -1\end{array}\right]\,\,.\end{equation}

Here as well, there is a correspondence with SU(2) generators, since
\begin{equation}
\widehat{\sigma}^{\pm}=\frac{1}{2}\left(\widehat{\sigma}^{x}\pm i\widehat{\sigma}^{x}\right),\end{equation}
 and \begin{equation}
\widehat{\sigma}^{z}=\hat{R}^{11}-\hat{R}^{00}.\end{equation}
 Identities can either be obtained from these correspondences, or
from direct differentiation, since:\begin{equation}
\frac{\partial}{\partial z}\left[\begin{array}{c}
e^{z/2}\\
e^{-z/2}\end{array}\right]=\frac{1}{2}\left[\begin{array}{c}
e^{z/2}\\
-e^{-z/2}\end{array}\right]\,\,.\end{equation}
 Hence, in operator language: \begin{equation}
\frac{\partial}{\partial z}\left\Vert z\right\rangle =\frac{1}{2}\widehat{\sigma}^{z}\left\Vert z\right\rangle \,\,.\label{eq:state-ident-1}\end{equation}
 On taking the hermitian transpose: \begin{equation}
\frac{\partial}{\partial z^{*}}\left\langle z\right\Vert =\frac{1}{2}\left\langle z\right\Vert \widehat{\sigma}^{z}\,\,.\label{eq:state-ident-2}\end{equation}

With a little algebra, one can also show that\begin{equation}
e^{\mp z}\left[\frac{1}{2}\pm\frac{\partial}{\partial z}\right]\left\Vert z\right\rangle =\widehat{\sigma}^{\pm}\left\Vert z\right\rangle \,\,.\end{equation}

\subsection{Equivalent Identities}

Here the functions differentiated are all analytic functions, either
of $z$ or of $z^{*}$. This means that we can always use Cauchy's
equivalence of differentiations in real and imaginary directions,
i.e., \begin{eqnarray}
\frac{\partial}{\partial z}\left\Vert z\right\rangle  & = & \frac{\partial}{\partial r}\left\Vert z\right\rangle =\frac{-i\partial}{\partial\phi}\left\Vert z\right\rangle \,\,,\nonumber \\
\frac{\partial}{\partial z^{*}}\left\langle z\right\Vert \, & = & \frac{\partial}{\partial r}\left\langle z\right\Vert =\frac{i\partial}{\partial\phi}\left\langle z\right\Vert \,\,.\end{eqnarray}
 This freedom, which also applies in the SU(n) case, allows one to
derive a variety of different equivalent equations for a given operator
evolution equation.

\section{SU(n) Phase-space }

Just as with the harmonic-oscillator coherent states, it is possible
to define a variety of operator representations using the SU($n$)
coherent states. A number of these have been extensively studied,
including representations analogous to the W\cite{Wig-Wigner}, Q\cite{Wig-Wigner},
P\cite{GSp}, and +P\cite{positiveP1,pp} representations. Spin
versions of the Q-representation\cite{Gilmore}, P-representation\cite{Narducci}
and Wigner representations\cite{ChatWig} have been introduced previously.
These essentially are defined on classical phase-spaces, in the sense
that the phase-space dimension is the same as that of the generators
of the coherent state.

However, as in the case of the harmonic oscillator, these do not generally
allow time-evolution equations with a stochastic (positive) propagator.
The difficulty here is that in general, these types of phase-space
representation do not give rise to a positive-definite diffusion and
hence to stochastic equations that can be numerically simulated.

Instead, we will focus on the SU(2) and SU(n) cases analogous to the
positive P representation\cite{positiveP1,pp}. This approach includes
off-diagonal projection operators in the expansion of the density
matrix, and give rise to a phase-space dimension which is at least
twice that of the classical phase-space. The result is a complete,
positive representation that generates positive-definite Fokker-Planck
equations. This generalizes related work in quantum and atom optics\cite{SmithGard,carusottothermo},
which uses similar procedures.

\subsection{SU(2) phase-space expansions}

We now illustrate these ideas with reference to the simplest SU(2)
or qubit case, using the reduced $z-$parametrization. If the density
matrix is separable, one can use a representation in terms of a positive
probability over the SU(2) diagonal coherent-state projectors:

\begin{equation}
{\widehat{\rho}}=\int P^{(2)}\left(\mathbf{z}\right)\left|\mathbf{z}\right\rangle \left\langle \mathbf{z}\right|d\mathbf{z}\,\,.\end{equation}
 It is always possible to define a positive representation like the
Husimi Q-function, which is:

\begin{equation}
Q^{(2)}\left(\mathbf{z}\right)=\left\langle \mathbf{z}\right|\widehat{\rho}\left|\mathbf{z}\right\rangle \,\,.\end{equation}

However, these two methods will not generally give a positive-definite
diffusion in the time-evolution equations for the distribution, except
in special cases. In order to achieve this, we must introduce off-diagonal
coherent state projectors, resulting in an expansion of form:

\begin{equation}
\widehat{\rho}=\int P^{(2)}\left(\lambda\right)\widehat{\Lambda}^{(2)}\left(\lambda\right)d\lambda\,\,.\end{equation}
 Here we define $\lambda=\left(\mathbf{z},\mathbf{z}'\right)$, so
that $d\lambda\equiv d^{2N}\mathbf{z}d^{2N}\mathbf{z}'$, and we have
introduced a general kernel operator with an arbitrary weight coefficient
$w$:\begin{equation}
\widehat{\Lambda}_{w}^{(2)}\left(\lambda\right)=\widehat{\Lambda}_{w}^{(2)}\left(\mathbf{z},\mathbf{z}'\right)=\left\Vert \mathbf{z}\right\rangle \left\langle \mathbf{z}'\right\Vert e^{w(\mathbf{z},\mathbf{z}')}\,\,.\end{equation}

With the simplest choice of $w=0$, we obtain an expansion in terms
of un-normalized projectors, which from Eq (\ref{eq:innerprod2})
leads to the result that

\begin{equation}
{\widehat{\rho}}=\int P^{(2)}\left(\mathbf{z},\mathbf{z}'\right)\widehat{\Lambda}_{0}^{(2)}\left(\mathbf{z},\mathbf{z}'\right)d^{2N}\mathbf{z}d^{2N}\mathbf{z}'\,\,,\end{equation}
 with a trace given by

\begin{eqnarray}
\Tr\left(\widehat{\Lambda}_{0}^{(2)}\left(\mathbf{z},\mathbf{z}'\right)\right) & = & \left\langle \mathbf{z}'\right.\left\Vert \mathbf{z}\right\rangle \nonumber \\
 & = & \prod_{j=1}^{N}\left[2\cosh\left(\left[z_{j}^{*}+z_{j}'\right]/2\right)\right]\nonumber \\
 & = & \Lambda\left(\mathbf{R}\right),\end{eqnarray}
 where we have introduced the kernel trace $\Lambda\left(\mathbf{R}\right)$
as a function of the combined variable $\mathbf{R}=\left[\mathbf{z}^{*}+\mathbf{z}'\right]/2$.

There are many other choices of weights and phase-space expansions.
One choice is to define the weight $w(\mathbf{z},\mathbf{z}')=-\ln\left\langle \mathbf{z}'\right.\left\Vert \mathbf{z}\right\rangle $.
This choice ensures that the kernel has a unit trace, giving results
analogous to the positive-P approach. In this case: \begin{equation}
\widehat{\Lambda}_{w}^{(2)}\left(\lambda\right)\equiv\widehat{\Lambda}_{+}^{(2)}\left(\mathbf{z},\mathbf{z}'\right)=\frac{\left\Vert \mathbf{z}\right\rangle \left\langle \mathbf{z}'\right\Vert }{\left\langle \mathbf{z}'\right.\left\Vert \mathbf{z}\right\rangle }\,\,.\end{equation}

More generally, either using $\lambda_{0}$ as a dynamical variable,
or other choices of weight function are necessary, in order to eliminate
boundary terms which can arise in dynamical equations\cite{GGD-Validity,GaugeP}.

\subsection{Entanglement and Bell states}

We note here that there is a fundamental contrast between this approach
and the diagonal P-representation approach originally due to Sudarshan
and Glauber\cite{GSp}, and later extended to SU(2) coherent states\cite{ACGT}.
The basis set of the diagonal P-representation is separable: it therefore
cannot represent entanglement, except as a limit of a generalized
function.

By comparison, the present approach includes terms that are fundamentally
inseparable, and therefore can represent states like Bell states.
To see this, consider the Bell state defined as: \begin{eqnarray}
\left|\psi^{B}\right\rangle  & = & \frac{1}{\sqrt{2}}\left[\left|0,1\right\rangle -\left|1,0\right\rangle \right]\nonumber \\
 & = & \frac{1}{\sqrt{2}}\left[\left|\underline{\overrightarrow{\psi}}^{+}\right\rangle -\left|\underline{\overrightarrow{\psi}}^{-}\right\rangle \right]\nonumber \\
 & = & \frac{1}{\sqrt{2}}\left[\left|\underline{\overrightarrow{\psi}}^{+}\right\rangle +\left|-\underline{\overrightarrow{\psi}}^{-}\right\rangle \right]\,\,.\end{eqnarray}

where:\begin{eqnarray}
\underline{\overrightarrow{\psi}}^{+} & = & \left[\begin{array}{cc}
1 & 0\\
0 & 1\end{array}\right]\nonumber \\
\underline{\overrightarrow{\psi}}^{-} & = & \left[\begin{array}{cc}
0 & 1\\
1 & 0\end{array}\right]\end{eqnarray}

The corresponding density matrix is:

\begin{eqnarray}
\widehat{\rho}^{B} & = & \frac{1}{2}\left[\left|\underline{\overrightarrow{\psi}}^{+}\right\rangle -\left|\underline{\overrightarrow{\psi}}^{-}\right\rangle \right]\left[\left\langle \underline{\overrightarrow{\psi}}^{+}\right|-\left\langle \underline{\overrightarrow{\psi}}^{-}\right|\right]\,\,.\nonumber \\
 & = & \frac{1}{2}\left[\left|\underline{\overrightarrow{\psi}}^{+}\right\rangle +\left|-\underline{\overrightarrow{\psi}}^{-}\right\rangle \right]\left[\left\langle \underline{\overrightarrow{\psi}}^{+}\right|+\left\langle -\underline{\overrightarrow{\psi}}^{-}\right|\right]\,\,.\end{eqnarray}

This has the form of a positive distribution over the off-diagonal
coherent state basis terms, as required.

\subsection{SU(n) phase-space expansions}

We now consider the most general SU(n) case. It is well-known\cite{Gilmore}
that one can define a diagonal phase-space representation analogous
to the Glauber P-function:

\begin{equation}
\widehat{\rho}=\int P^{(n)}\left(\underline{\overrightarrow{\psi}}\right)\left|\underline{\overrightarrow{\psi}}\right\rangle \left\langle \underline{\overrightarrow{\psi}}\right|d\underline{\overrightarrow{\psi}}\,\,.\end{equation}
 A positive Q-function like phase-space representation always exists,
with:

\begin{equation}
Q^{(n)}\left(\underline{\overrightarrow{\psi}}\right)=\left\langle \underline{\overrightarrow{\psi}}\right|\widehat{\rho}\left|\underline{\overrightarrow{\psi}}\right\rangle \,\,.\end{equation}

Just as in the SU($2$) case, neither of these phase-space methods
will usually result in positive-definite stochastic evolution, either
for canonical ensembles or for dynamical evolution. To overcome this
limitation, a positive representation using off-diagonal projectors
must be introduced:

\begin{equation}
{\widehat{\rho}}=\int P^{(n)}\left(\lambda\right)\widehat{\Lambda_{w}}^{(n)}\left(\lambda\right)d\lambda\,\,.\end{equation}
 Here we define $\lambda=\left(\lambda_{0,}\underline{\overrightarrow{\psi},}\underline{\overrightarrow{\phi}}\right)$,
so that $d\lambda\equiv d^{2(d+1)}\lambda=d^{2}\lambda_{0}d^{2Mn}\underline{\overrightarrow{\psi}}d^{2Mn}\underline{\overrightarrow{\phi}}$
where $d=2Mn,$ together with a general kernel operator with weight
coefficient $w$:\begin{equation}
\widehat{\Lambda}_{w}^{(n)}\left(\lambda\right)={\widehat{\Lambda}_{w}^{(n)}}\left(\underline{\overrightarrow{\psi},}\underline{\overrightarrow{\phi}}\right)=\left\Vert \mathbf{\underline{\overrightarrow{\psi}}}\right\rangle ^{(n,M)}\left\langle \underline{\overrightarrow{\phi}}\right\Vert ^{(n,M)}e^{\lambda_{0}+w(\underline{\overrightarrow{\psi}},\underline{\overrightarrow{\phi}})}\,\,.\end{equation}

This reduces to the diagonal case when $\underline{\overrightarrow{\psi}}=\underline{\overrightarrow{\phi}}$.
From Eq (\ref{eq:innerprod2}), the simplest choice of $\lambda_{0}=w=0$
leads to the result that:

\begin{eqnarray}
\Tr\left({\widehat{\Lambda}_{0}^{(n)}}\left(\underline{\overrightarrow{\psi},}\underline{\overrightarrow{\phi}}\right)\right) & = & \left\langle \underline{\overrightarrow{\phi}}\right\Vert ^{(n,M)}\left\Vert \mathbf{\underline{\overrightarrow{\psi}}}\right\rangle ^{(n,M)}\nonumber \\
 & = & \prod_{m=1}^{M}\left[\overrightarrow{\phi_{m}^{*}}\cdot\overrightarrow{\psi}{}_{m}\right]^{N}\nonumber \\
 & = & \Lambda^{(n)}\left(\underline{\overrightarrow{\psi},}\underline{\overrightarrow{\phi}}\right),\end{eqnarray}

Another choice is to define the weight \begin{equation}
w\left(\underline{\overrightarrow{\psi},}\underline{\overrightarrow{\phi}}\right)=-\ln\left\langle \underline{\overrightarrow{\phi}}\right\Vert \left\Vert \mathbf{\underline{\overrightarrow{\psi}}}\right\rangle \,,\end{equation}
 so that the kernel has a unit trace, giving results analogous to
the positive-P approach. However, unless there is damping, this choice
by itself can lead to instabilities and boundary term errors\cite{GGD-Validity}.

If $\lambda_{0}\neq0$, it can be used as another dynamical variable,
giving stabilized weighted trajectories as in the stochastic gauge
method\cite{GaugeP}. More general weight choices are also possible.
The use of different weights changes the form of the resulting dynamical
equations, thereby giving rise to useful techniques which can be utilized
to optimize and solve these equations. An example will be given in
the next section.

\section{Dynamical calculations}

The calculation of observables and correlations in real or imaginary
time (for thermal equilibrium) is the main purpose of this phase-space
method. The advantage of the approach is that it is a general-purpose
method. The identities and transformations involved do not depend
on detailed properties of the Hamiltonian, apart from the requirement
that it must be able to be expressed using the group generators.

Provided this requirement is satisfied, the calculations involved
are not specific to a given model. However, some caution is necessary.
The probability distributions obtained can have a variety of widths
in phase-space, which means there is a large range of potential sampling
errors possible. This is not uniquely specified by the Hamiltonian.
As the SU($n$) basis set is not orthogonal, the phase-space distribution
is therefore not unique, and depends on the precise identities and
algorithms chosen. Since the underlying coherent states factorize
on a lattice, one may expect that increasing correlations and entanglement
between lattice sites will require an increased `footprint' of the
distribution, and hence an increased sampling error.

\subsection{General evolution problems}

To illustrate the procedure, the required dynamical evolution is first
written as a Liouville equation for the density operator. This may
or may not be unitary, and does not have to be trace-preserving, as
long as it is linear in $\hat{\rho}$, and can be written using a
polynomial in the group generators: \begin{eqnarray}
\partial\widehat{\rho}(t)/\partial t & = & \widehat{L}[\widehat{\rho}(t)]\,.\end{eqnarray}
 To solve this with phase-space methods, we first expand the density
operator over the SU(n) operator basis $\widehat{\Lambda}^{(n,M)}(\overrightarrow{\lambda})$,
where $\overrightarrow{\lambda}$ is the set of all complex coherent
amplitudes:\begin{eqnarray}
\widehat{\rho}(t) & = & \int P(\overrightarrow{\lambda},t)\widehat{\Lambda}(\overrightarrow{\lambda})d\overrightarrow{\lambda}.\end{eqnarray}
 This defines a $d+1$-dimensional complex phase-space, where $d=2Mn$
as before, with a dynamical weight variable $\lambda_{0}$ if necessary.
The SU(n) differential identities allow us to write the Liouville
operator equation as\begin{eqnarray}
\partial\widehat{\rho}(t)/\partial t & = & \int P(\overrightarrow{\lambda},t)\mathcal{L}_{A}\widehat{\Lambda}(\overrightarrow{\lambda})d^{2(d+1)}\overrightarrow{\lambda},\end{eqnarray}
 where $\mathcal{L}_{A}$ is a linear differential operator. Due to
the non-uniqueness of the identities, this can include arbitrary \emph{stochastic
gauge} functions. Provided there are no derivatives higher than second
order, this equation can finally be transformed into a positive-definite,
weighted Fokker-Planck equation for $P$. It is essential that the
gauges are chosen to eliminate any boundary terms that may otherwise
arise from the partial integration\cite{GaugeP,GGD-Validity}.\begin{equation}
\frac{\partial}{\partial t}P(\overrightarrow{\lambda},t)=\left[U-\frac{\partial}{\partial\lambda_{\mu}}A_{\mu}+\frac{1}{2}\frac{\partial^{2}}{\partial\lambda_{\mu}\partial\lambda_{\nu}}D_{\mu\nu}\right]P(\overrightarrow{\lambda},t)\,.\end{equation}

Here we use a summation convention where $\mu=1,d$. Introducing a
matrix square root $B$, where $D_{\mu\nu}=B_{\mu\rho}B_{\nu\rho}$,
this can then be transformed into the stochastic equations, which
in Ito calculus are generically of the form:\begin{eqnarray}
d\lambda_{0}/\partial t & = & U+g_{\mu}\zeta_{\mu}-\frac{1}{2}g_{\mu}g_{\mu}\nonumber \\
d\lambda_{\mu}/\partial t & = & A_{\mu}+B_{\mu\nu}\left(\zeta_{\mu}-g_{\mu}\right).\label{eq:gauge2}\end{eqnarray}
 Here the weight term $U$ and the drift vector $\mathbf{A}$ are
determined by the form of the original Liouville equation. The drift
gauges appear as the arbitrarily functions $\mathbf{g}$, and diffusion
gauges appear as the freedom that exists in choosing the noise matrix
$\mathbf{B}$. The noise terms $\bm{\zeta}$ are Gaussian white noises,
with correlations:\begin{equation}
\langle\zeta_{\mu}(t)\zeta_{\nu}(t')\rangle=\delta_{\mu\nu}\delta(t-t').\label{eq:gauge3}\end{equation}

Equations (\ref{eq:gauge2}-\ref{eq:gauge3}) can be used to solve
a large class of quantum dynamical and thermal-equilibrium problems
in coherent-state representations. In practice, the numerical implementation
of these equations can be simplified by use of automatic code-generators\cite{Collecut,xmds}.

\subsection{Operator identities: SU(2) case}

To use this approach, one must obtain differential identities for
the group generators. We start with the SU($2$) case. Here we will
omit the superscript $(2)$ indicating an SU(2) kernel, when there
is no ambiguity.

With the simplest constant weight choice we will use here of $w=0$,
the only differential identities needed are obtained directly from
Eq (\ref{eq:state-ident-1}) and Eq (\ref{eq:state-ident-2}) i.e.,

\begin{eqnarray}
\frac{\partial}{\partial z}\widehat{\Lambda}_{0} & = & S^{z}\widehat{\Lambda}_{0},\nonumber \\
\frac{\partial}{\partial z'*}\widehat{\Lambda}_{0} & = & \widehat{\Lambda}_{0}\widehat{S}^{z}.\end{eqnarray}
 Other useful differential identities in more general cases are\begin{eqnarray}
\frac{\partial}{\partial z}{\widehat{\Lambda}_{w}} & = & \left[S^{z}+\frac{\partial w}{\partial z}\right]{\widehat{\Lambda}_{w}},\nonumber \\
\frac{\partial}{\partial z'*}{\widehat{\Lambda}_{w}} & = & {\widehat{\Lambda}_{w}}\left[\widehat{S}^{z}+\frac{\partial w}{\partial z'*}\right]\,.\end{eqnarray}

Hence, for example, one can write:\begin{eqnarray}
\widehat{S}^{z}{\widehat{\Lambda}_{w}} & = & \left[\frac{\partial}{\partial z}-\frac{\partial w}{\partial z}\right]{\widehat{\Lambda}_{w}},\\
{\widehat{\Lambda}_{w}}\widehat{S}^{z} & = & \left[\frac{\partial}{\partial z'*}-\frac{\partial w}{\partial z'*}\right]{\widehat{\Lambda}_{w}}.\nonumber \end{eqnarray}

\subsection{Operator identities: SU(n) case}

We wish to obtain similar differential identities for the SU(n) coherent
state kernels. These are:\begin{eqnarray}
\psi_{m}^{\nu}\frac{\partial}{\partial\psi_{m}^{\mu}}\widehat{\Lambda}_{w}^{(n)} & = & \left[\hat{R}_{m}^{\mu\nu}+\psi_{m}^{\nu}\frac{\partial w}{\partial z}\right]\widehat{\Lambda}_{w}^{(n)},\nonumber \\
\phi_{m}^{\nu*}\frac{\partial}{\partial\phi_{m}^{\mu*}}\widehat{\Lambda}_{w}^{(n)} & = & \widehat{\Lambda}_{w}^{(n)}\left[\hat{R}_{m}^{\nu\mu}+\phi{}_{m}^{\nu*}\frac{\partial w}{\partial\psi\phi{}_{m}^{\mu*}}\right]\,.\end{eqnarray}

Since each occurrence of a group generator $\hat{R}$ gives rise to
a differential term, the requirement that time-evolution is stochastic
corresponds to a restriction to Hamiltonians and master equations
that are quadratic in the group generators or generalized spin operators.

\subsection{Observables}

We illustrate how to calculate observables by reference to the the
spin-half system , where the main observable of interest is the magnetization
at site $i$, given by:\begin{equation}
\left\langle \widehat{\sigma}_{i}^{z}\right\rangle =\frac{Tr\left(\widehat{\sigma}_{i}^{z}{\widehat{\rho}}\right)}{Tr\left({\widehat{\rho}}\right)}\,\,.\end{equation}

Defining the normalization as $Z=Tr\left({\widehat{\rho}}\right)$,
with a measure $d\lambda=d^{2N}\mathbf{z}d^{2N}\mathbf{z}'$, one
obtains that the uniform weight expansion case has the normalization
\begin{eqnarray}
Z & = & \int P\left(\mathbf{z},\mathbf{z}'\right)\Lambda\left(\mathbf{R}\right)d\lambda\,\,,\nonumber \\
 & = & \left\langle \Lambda\left(\mathbf{R}\right)\right\rangle _{P}.\end{eqnarray}
 Noting that \begin{equation}
\Tr\left(\widehat{\sigma}_{i}^{z}\widehat{\Lambda}_{0}\right)=\tanh\left(R_{i}\right)\prod_{j=1}^{N}\left[2\cosh\left(R_{j}\right)\right]\,\,,\end{equation}
 we can introduce a c-number equivalent magnetization variable $m_{j}=\tanh\left(R_{j}\right)$.
The mean magnetization is then written as \begin{eqnarray}
\left\langle \widehat{\sigma}_{i}^{z}\right\rangle  & = & \int P\left(\mathbf{z},\mathbf{z}'\right)\tanh\left(R_{i}\right)\Lambda\left(\mathbf{R}\right)d\lambda\nonumber \\
 & = & \frac{1}{Z}\left\langle \tanh\left(R_{i}\right)\Lambda\left(\mathbf{R}\right)\right\rangle _{P}\,.\end{eqnarray}

Similarly, the correlation function between two different sites is
\begin{equation}
\left\langle \widehat{\sigma}_{i}^{z}\widehat{\sigma}_{j}^{z}\right\rangle =\frac{1}{Z}\left\langle \tanh\left(R_{i}\right)\tanh\left(R_{j}\right)\Lambda\left(\mathbf{R}\right)\right\rangle _{P}\,\,.\end{equation}

\subsection{Phase-independent case}

In the case where the Hamiltonian is only a function of $\widehat{\sigma}^{z}$'s
-- as in the Ising model, considered in the next section -- a much
simpler expansion of the density operator can be used. While this
is less general, it provides an alternative way to derive the results
in the next section.

This simplified expansion is: \begin{equation}
\widehat{\rho}=\int P(\mathbf{R})\widehat{\Lambda}_{z}\left(\mathbf{R}\right)d\mathbf{R}\,,\end{equation}
 where $\widehat{\Lambda}_{z}\left(\mathbf{R}\right)$ is obtained
on phase-averaging over the complete kernel, with the result that:
\begin{eqnarray}
\widehat{\Lambda}_{z}\left(\mathbf{R}\right) & = & \prod_{j=1}^{M}\exp(R_{j}\widehat{\sigma}_{j}^{z})\nonumber \\
 & = & \prod_{j=1}^{M}\sum_{n=0}^{\infty}\frac{(R_{j}\widehat{\sigma}_{j}^{z})}{n!}\nonumber \\
 & = & \prod_{j=1}^{M}2\left(\cosh(R_{j})+\widehat{\sigma}_{j}^{z}\sinh(R_{j})\right).\end{eqnarray}
 The operator correspondence \begin{equation}
\widehat{\sigma}_{j}^{z}\widehat{\Lambda}_{z}\left(\mathbf{R}\right)=\frac{\partial}{\partial R_{j}}\widehat{\Lambda}_{z}\left(\mathbf{R}\right)\end{equation}
 then holds.

In the following section, we will focus on using the full coherent
state identities, as these are more generally applicable. However,
we note that for those primarily interested in the Ising model, our
results can also be readily obtained using this reduced expansion.

\section{The Ising model}

As an instructive example, we show that a lattice of SU(2) coherent
states can be used to solve for the partition function of the Ising
model of interacting spins. This is the simplest nontrivial case where
one obtains an exactly soluble phase-transition in a spin model in
two dimensions. As well having a wide applicability, it does illustrate
many of the fundamental scaling issues that occur in using phase-space
methods to solve coupled spin models. Similar features also occur
in more complex quantum spin models, which will be treated in greater
detail elsewhere.

Although we focus here on the simplest case possible where $S=1/2$
at each site, we note that the basic ideas also hold for more general
coupled spin-$S$ spin systems, or interacting atoms described by
the most general $SU(n)$ coherent states. However, in this example
we make use of some identities and simplifying features that are unique
to the spin-half case.

The most general form of this model -- in a summation convention which
sums repeated indices -- has the Hamiltonian\begin{equation}
{\widehat{H}}=-\sum_{i}h_{i}\widehat{\sigma}_{i}^{z}-\frac{1}{2}\sum_{ij}J_{ij}\widehat{\sigma}_{i}^{z}\widehat{\sigma}_{j}^{z}.\end{equation}

We will assume here that the coupling term $J_{ij}$ is symmetric,
with $J_{ij}\geq0$, which corresponds to attractive interactions
between spins. Since $\left(\widehat{\sigma}_{j}^{z}\right)^{2}=1$,
self-interactions have no effect apart from shifting the energy origin,
and therefore it is common to set $J_{ii}=0$ for simplicity. Different
choices of $J_{ij}$ will generate different types of Ising model,
that can have any dimensionality, shape, or distribution of interaction
strengths. The choice of $J_{ij}=J$ for all nearest neighbours corresponds
to the standard Onsager model\cite{Onsager}. The interaction terms
will be called links, since they typically join neighbouring spin
sites or nodes on a lattice.The factor of half in the Hamiltonian
accounts for the fact that all links are counted twice in the double
summation.

The density matrix, which gives information about the spin distribution
in thermal equilibrium, is \begin{equation}
{\widehat{\rho}}=\exp\left(-\beta{\widehat{H}}\right).\end{equation}

One often wishes to calculate the total partition function $Z$, where
\begin{equation}
Z=\Tr\left({\widehat{\rho}}\right).\end{equation}
 If all the terms $J_{ij}$ are either equal to each other or zero,
then the interactions are uniquely characterized by a graph showing
which nodes are linked by a nonzero interaction. Hence, there is a
close relationship between the Ising model, and mathematical problems
that count paths on a lattice. Once the total number of ways of constructing
links with a given energy is known, the partition function can be
easily obtained. Since there are $2^{M}$ distinct spin configurations,
it is exponentially difficult to evaluate this directly, unless special
types of symmetry occur which can sometimes lead to exact solutions.
Examples are the case of the one and two dimensional regular lattice
with uniform nearest-neighbour interactions, and the simplex with
all node-pairs linked equally.

More generally, one must use probabilistic methods to sample the spin
configurations. The standard techniques involve Monte Carlo or Metropolis
techniques in which spins are flipped randomly, in order to obtain
an ensemble of spin configurations at a fixed temperature. There is
a long history to these methods, which can give excellent results.
However, while much more efficient than direct configuration counting,
these methods are still computationally intensive. This means that
there are often strong limits to either the size of the lattice or
to the accuracy, which is limited by the sampling error. Recent improvements
in these standard techniques involve flipping clusters of spins, which
is more effective at the critical temperature where the correlation
lengths are large.

We consider a different approach to this calculation using a differential
equation method, that uses the atomic coherent state basis with a
continuous parameter, rather than discrete spin configurations. The
density operator satisfies the following equation\cite{DDK2004}:
\begin{equation}
\frac{\partial\widehat{\rho}\left(\beta\right)}{\partial\beta}=-\frac{1}{2}\left[\widehat{\rho}\left(\beta\right),{\widehat{H}}\right]_{+}.\end{equation}

The initial condition at high temperature is just \begin{equation}
\widehat{\rho}\left(0\right)=\hat{1}=\otimes_{j=1}^{M}\hat{1}_{j}.\end{equation}

\subsection{Fokker-Planck Equation}

Next, the partition function ${\widehat{\rho}}$ is expanded using
an SU(2) coherent state projector basis, so that \begin{equation}
\widehat{\rho}\left(\beta\right)=\int P\left(\mathbf{z},\mathbf{z}',\beta\right)\widehat{\Lambda}\left(\mathbf{z},\mathbf{z}'\right)d\lambda\,\,.\end{equation}
 From the two-level completeness identity, Eq (\ref{eq:2-level-completeness}),
one can write \begin{equation}
P\left(\mathbf{z},\mathbf{z}',0\right)=\prod_{i=1}^{M}\left[\frac{1}{2\pi}\delta\left(\theta_{i}-\theta_{i}'\right)\delta\left(r_{i}\right)\delta\left(r_{i}'\right)\right].\end{equation}
 This involves a single unique $r$ value, $r_{j}=0$, and a random
phase. This is transformed using operator identities into the resulting
Fokker-Planck equation is transformed to a stochastic differential
equation that can be sampled. We can choose equations in which the
initially random phase is invariant. This leads to a stochastic equation
in $r_{j}$ in which the initial state is given exactly, without sampling
error. This technique can also be written as a type of path-integral.

To illustrate the idea, we start with the simplest unweighted kernel,
as previously:\begin{eqnarray}
\widehat{\rho}\left(\beta\right) & = & \int P\left(\lambda,\beta\right)\widehat{\Lambda}\left(\lambda\right)d\lambda\nonumber \\
 & = & \int P\left(\mathbf{z},\mathbf{z}',\beta\right)\left\Vert \mathbf{z}\right\rangle \left\langle \mathbf{z}'\right\Vert d\lambda.\end{eqnarray}

We see from this that\begin{equation}
\frac{\partial\widehat{\rho}\left(\beta\right)}{\partial\beta}=-\frac{1}{2}\int P\left(\lambda,\beta\right)\left[{\widehat{\Lambda}}\left(\lambda\right),{\widehat{H}}\right]_{+}d\lambda.\end{equation}

Introducing the mean interaction strength per spin, \begin{equation}
\bar{J}=\frac{1}{2N}\sum_{i,j}J_{ij},\end{equation}
 we then rewrite the Hamiltonian in a form that allows us to obtain
positive-definite diffusion terms, \begin{eqnarray}
{\widehat{H}} & = & -h_{i}\widehat{\sigma}_{i}^{z}-\frac{1}{4}J_{ij}\left(\widehat{\sigma}_{i}^{z}+\widehat{\sigma}_{j}^{z}\right)^{2}+N\bar{J}\nonumber \\
 & = & {\widehat{H}}'+M\bar{J}.\end{eqnarray}

The constant term has no effect on observable quantities, and will
be neglected in the following calculations. In other words, we will
calculate \begin{equation}
{\widehat{\rho}}'=\exp\left(-\beta{\widehat{H}}'\right)={\widehat{\rho}}e^{\beta M\bar{J}},\end{equation}
 which differs from the ${\widehat{\rho}}$ defined above only by
an overall normalization factor. Inserting the relevant identities,
the two different operator orderings give \begin{eqnarray}
-\frac{1}{2}{\widehat{H}}'{\widehat{\Lambda}} & = & \frac{1}{2}\left[h_{i}\widehat{\sigma}_{i}^{z}+\frac{1}{4}J_{ij}\left(\widehat{\sigma}_{i}^{z}+\widehat{\sigma}_{j}^{z}\right)^{2}\right]{\widehat{\Lambda}}\nonumber \\
 & = & \left[h_{i}\partial_{i}+\frac{1}{2}J_{ij}\left(\partial_{i}+\partial_{j}\right)^{2}\right]{\widehat{\Lambda}},\end{eqnarray}
 and: \begin{eqnarray}
-\frac{1}{2}{\widehat{\Lambda}}{\widehat{H}}' & = & \frac{1}{2}{\widehat{\Lambda}}\left[h_{i}\widehat{\sigma}_{i}^{z}+\frac{1}{4}J_{ij}\left(\widehat{\sigma}_{i}^{z}+\widehat{\sigma}_{j}^{z}\right)^{2}\right]\nonumber \\
 & = & \left[h_{i}{\partial'}_{i}+\frac{1}{2}J_{ij}\left({\partial'}_{i}+{\partial'}_{j}\right)^{2}\right]{\widehat{\Lambda}}.\end{eqnarray}

Here we have used the definitions $\partial_{i}\equiv\partial/\partial r_{i}$
and ${\partial'}_{i}\equiv\partial/\partial{r'}_{i}$. We now introduce
an extended vector notation with indices $\mu=1,\ldots,2N$, so that
$r_{N+j}\equiv r_{j}'$ and $\partial_{\mu}=\partial/\partial r_{\mu}$
, with coupling constants $J_{\mu\nu}$, $h_{\mu}$ defined so that
$J_{i+N,j+N}=J_{ij}$, and $h_{i+N}=h_{i}$ .

Next, on integrating by parts, and equating coefficients of $\widehat{\Lambda}$,
one obtains the following Fokker-Planck equation, with explicitly
positive definite diffusion terms:\begin{equation}
\frac{\partial P}{\partial\beta}=\left[-\sum_{\mu}h_{\mu}\partial_{\mu}+\frac{1}{2}\sum_{\mu\nu}J_{\mu\nu}\left(\partial_{\mu}+\partial_{\nu}\right)^{2}\right]P\,\,.\end{equation}

\subsection{Stochastic Equation}

To obtain an equivalent stochastic equation, we must first write the
Fokker-Planck equation in the form:

\begin{equation}
\frac{\partial P}{\partial\beta}=\partial_{\mu}\left[-A_{\mu}+\frac{1}{2}D_{\mu\nu}\partial_{\nu}\right]P\,\,,\end{equation}
 A suitable factorized diffusion matrix form is readily found by expanding
the diffusion matrix $D_{\mu\nu}$ as a sum over distinct terms for
each non-vanishing link, that is:\begin{equation}
\underline{\underline{D}}=\sum_{\mu,\nu}^{2N}J_{\mu\nu}\left[\begin{array}{c}
\vdots\\
1_{\mu}\\
\vdots\\
1_{\nu}\\
\vdots\end{array}\right]\left[\ldots1_{\mu},\ldots,1_{\nu},\ldots\right]\,.\end{equation}

It is immediate that $\mathbf{D}$ can be factorized in the form:

\begin{equation}
\underline{\underline{D}}=\sum_{\mu,\nu}^{2N}J_{\mu\nu}\mathbf{B}^{(\mu\nu)}\mathbf{B}^{(\mu\nu)T}\,\,,\end{equation}
 where $\mathbf{B}^{(\mu\nu)}$ is a $2N$ dimensional vector with
two non-vanishing entries at $\mu$ and $\nu$ respectively, i.e.,

\begin{equation}
\mathbf{B}^{(\mu\nu)}=\left[\begin{array}{c}
\vdots\\
1_{\mu}\\
\vdots\\
1_{\nu}\\
\vdots\end{array}\right]\,\,.\end{equation}
 The corresponding stochastic equations are then:\begin{eqnarray}
\frac{\partial r_{\mu}}{\partial\beta} & = & A_{\mu}+\sum_{\mu',\nu'}^{2N}B_{\mu}^{(\mu'\nu')}\zeta_{\mu'\nu'}\nonumber \\
 & = & h_{\mu}+\sum_{\nu}\left(\zeta_{\mu\nu}+\zeta_{\nu\mu}\right)\,.\end{eqnarray}
 where the independent real stochastic noises $\zeta_{\mu\nu}$ are
correlated as \begin{equation}
\left\langle \zeta_{\mu\nu}\left(\beta\right)\zeta_{\mu'\nu'}\left(\beta'\right)\right\rangle =J_{\mu\nu}\delta_{\mu\mu'}\delta_{\nu\nu'}\delta\left(\beta-\beta'\right)\:.\end{equation}

These equations have the feature that they involve noise terms that
are automatically correlated between pairs of spins linked by an interaction
term, $J_{ij}$. The initial random phase is not changed by the interactions,
and only the magnetization -- which depends on $r_{j}$ -- changes
randomly in time. Spins that are linked tend to change together, as
they experience a correlated noise term.

Only the sum of $r_{j}+r_{j}'=2R_{j}$ is relevant to the observed
spin orientation. Defining \begin{equation}
W_{ij}\left(\beta\right)=\frac{1}{2}\int_{0}^{\beta}\left(\zeta_{ij}\left(\beta'\right)+\zeta_{i+N,j+N}\left(\beta'\right)\right)d\beta'\,\,,\end{equation}
 the resulting noise terms have a variance proportional to the inverse
temperature: \begin{equation}
\left\langle W_{\mu\nu}^{2}\left(\beta\right)\right\rangle =\frac{\beta J_{ij}}{2}\,\,.\end{equation}

\subsection{Partition function}

The solution at inverse temperature $\beta$ is:\begin{equation}
R_{i}\left(\beta\right)=h_{i}\beta+\sum_{j}W_{ij}^{+}\left(\beta\right)\,,\end{equation}
 where $W_{ij}^{+}\left(\beta\right)\equiv W_{ij}\left(\beta\right)+W_{ji}\left(\beta\right)$.
The resulting partition function is simply obtained on averaging over
all the stochastic trajectories, so that:\begin{eqnarray}
Z\left(\beta\right) & = & \left\langle \Lambda\left(\mathbf{R}\left(\beta\right)\right)\right\rangle e^{-\beta N\bar{J}}\nonumber \\
 & = & \left\langle \prod_{i}\left[2\cosh\left(R_{i}\left(\beta\right)\right)\right]\right\rangle e^{-\beta N\bar{J}}\,.\end{eqnarray}

This gives an explicit solution for the partition function as an expectation
value over the random processes $\mathbf{R}(\beta)$. We note that
while one may try to evaluate the partition function by simply averaging
over many stochastic trajectories, this is far from being an efficient
procedure. The problem is that the weights $\Lambda\left(\mathbf{R}\left(\beta\right)\right)$
grow exponentially large for large values of $\left|R_{j}\right|$,
which results in a large dispersion of trajectory weights, and therefore
extremely large sampling errors. This naive method is not practical.
A much more efficient procedure will be given in the next section.

We notice at this stage, however, an interesting feature of these
results. This is that the noise terms act only to couple adjacent
sites together. Thus, an understanding of the renormalization behaviour
of this problem can be realized by grouping spins together into clusters,
in which case the residual noise from cluster interactions scales
proportionate to the surface area of the cluster, rather than from
the total volume.

\subsubsection{Example: 2-site problem }

As an example of the simplest nontrivial case with a uniform external
field (i.e., $h_{i}=h$) the two-node partition function has only
one link, so \begin{equation}
{\widehat{H}}=-h\left(\widehat{\sigma}_{1}^{z}+\widehat{\sigma}_{2}^{z}\right)-J\widehat{\sigma}_{1}^{z}\widehat{\sigma}_{2}^{z}\,\,.\end{equation}
 There are four distinct states with interaction energies of $\pm J$.
Taking the trace, one can directly check from expanding over the four-dimensional
configuration space, that \begin{eqnarray}
Z_{2} & = & \Tr\left(e^{-\beta{\widehat{H}}}\right)\nonumber \\
 & = & e^{\beta\left(2h+J\right)}+2e^{-J\beta}+e^{\beta\left(-2h+J\right)}\,\,.\end{eqnarray}

For $h=0$, the two-site correlation can be calculated immediately
to be \begin{eqnarray}
\left\langle \widehat{\sigma}_{1}^{z}\widehat{\sigma}_{2}^{z}\right\rangle  & = & \frac{1}{Z\left(\beta\right)}Tr\left(\widehat{\sigma}_{1}^{z}\widehat{\sigma}_{2}^{z}e^{-\beta{\widehat{H}}}\right)\nonumber \\
 & = & \tanh\left(\beta J\right)\,.\label{eq:exact_two-site}\end{eqnarray}

We now wish to demonstrate how identical results are obtainable from
the raw stochastic equations. Introducing $W^{+}\left(\beta\right)=W_{12}\left(\beta\right)+W_{21}\left(\beta\right)$,
with $\left\langle W^{+2}\left(\beta\right)\right\rangle =\beta J$,
one finds that the two SU(2) coherent state amplitudes are always
equal to each other: \begin{equation}
R_{1}\left(\beta\right)=R_{2}\left(\beta\right)=h\beta+W^{+}\left(\beta\right)\,\,.\end{equation}
 Hence with $\bar{J}=J/2$, the partition function calculated from
the stochastic equations is \begin{eqnarray}
Z\left(\beta\right) & = & \left\langle \prod_{i}\left[2\cosh\left(R_{i}\left(\beta\right)\right)\right]\right\rangle _{P}e^{-\beta N\bar{J}}\nonumber \\
 & = & \left\langle \left[e^{R\left(\beta\right)}+e^{-R\left(\beta\right)}\right]^{2}\right\rangle _{P}e^{-\beta J}\,\,.\end{eqnarray}

Now, for a Gaussian process, \begin{equation}
\left\langle e^{\pm2R\left(\beta\right)}\right\rangle _{P}=\exp\left[\pm2h\beta+2\left\langle W^{2}\left(\beta\right)\right\rangle _{P}\right]=\exp\left[\pm2h\beta+2J\beta\right]\,\,,\end{equation}
 so the final result for the partition function is\begin{equation}
Z\left(\beta\right)=e^{\beta\left(2h+J\right)}+2e^{-J\beta}+e^{\beta\left(-2h+J\right)}\,\,.\end{equation}
 Similarly, for the correlation function in the limit of $h=0$:

\begin{eqnarray}
\left\langle \widehat{\sigma}_{1}^{z}\widehat{\sigma}_{2}^{z}\right\rangle  & = & \frac{e^{-\beta J}}{Z\left(\beta\right)}\left\langle \tanh^{2}\left(R\right)\Lambda\left(\mathbf{R}\right)\right\rangle _{P}\,\nonumber \\
 & = & \frac{4e^{-\beta J}}{Z\left(\beta\right)}\left\langle \sinh^{2}\left(R\right)\right\rangle _{P}\nonumber \\
 & = & \tanh\left(\beta J\right).\label{eq:exact-two-site-stoch}\end{eqnarray}

This agrees with the result from the direct calculation.

\section{Computational Strategies}

There are several possible strategies for calculating the partition
function while taking account of the final weight. For the Ising model,
a direct solution to the original stochastic equation is inefficient
for large $M$, as almost all trajectories will have an exponentially
small weight compared to a very small number of optimal trajectories.
We will demonstrate a strategy for making use of the fact that we
now have a solution to the stochastic equations in closed form, which
allows the problem to be re-sampled in a more efficient way.

\subsection{Optimized stochastic methods}

One way to solve this problem is to use weighted kernels or gauge
equations, combined with a strategy for breeding trajectories of largest
weight, which is essentially the diffusion Monte-Carlo approach\cite{Binder}.
Another approach is to use the Metropolis method\cite{Metropolis},
in which the link noise $W_{ij}$ is repeatedly randomized, based
on the final weight it generates, with some choices being accepted
and some being rejected.

A third way is to define a new stochastic equation whose solution
gives the link noise distribution, \emph{without} any additional weight.
To see this more clearly, suppose we write the final partition function
as a multi-component integral over the link noises $\mathbf{W}$,
including the Gaussian weight factor used to generate the noises $W_{ij}$:

\begin{equation}
Z\left(\beta\right)=\int\ldots\int d\mathbf{W}\exp\left(-V\left(\mathbf{W},\beta\right)\right)\,\,,\end{equation}
 where we have ignored all irrelevant normalization terms, and introduced
a potential that already includes the weight factor: \begin{equation}
V\left(\mathbf{W},\beta\right)=\sum_{i,j}\frac{1}{J_{ij}\beta}W_{ij}^{2}-\sum_{i}\ln\cosh\left(h_{i}\beta+\sum_{j}W_{ij}^{+}\left(\beta\right)\right)\,\,.\end{equation}

The first term is the most important at high temperatures. It tends
to keep all link noises small, so that the magnetization is nearly
zero. The second term is increasingly important at large $\beta$,
as it gives an increasing weight to terms with large correlated noises
$W_{ij}$, in which all links leading to a given spin have an identical
sign. This leads to formation of magnetized clusters.

A general Fokker-Planck equation that leads to the asymptotic solution
$\exp\left(-V\left(\mathbf{W},\beta\right)\right)$ at $\tau\rightarrow\infty$,
has the form:\begin{equation}
\frac{\partial{\cal P}}{\partial\tau}=\frac{1}{2}\partial_{\mathbf{i}}\left\{ {\cal D}_{\mathbf{i}\mathbf{j}}\left[\frac{\partial V}{\partial W_{\mathbf{j}}}+\partial_{\mathbf{j}}\right]\right\} {\cal P}\,\,,\end{equation}
 where we define $\mathbf{i}=\{i,j\}$, and differential operators
$\partial_{\mathbf{i}}\equiv\partial/\partial W_{\mathbf{i}}$. Differentiating
the potential $V$, one obtains:\begin{equation}
\frac{\partial V}{\partial W_{ij}}=\frac{2W_{ij}}{J_{ij}\beta}-\tanh\left(R_{i}\right)-\tanh\left(R_{j}\right)\,.\end{equation}

A range of stochastic equations for the link noises can be obtained,
by choosing different forms of the new diffusion matrix ${\cal D}_{\mathbf{j}\mathbf{j}'}$.
In particular, we note that one may expect that a diffusion matrix
that couples sites together over a distance of order of the expected
correlation length might be expected to give a particularly efficient
algorithm, as it tends to flip clusters of spins all of which have
a similar spin orientation. For simplicity, we do not investigate
this here, as we are interested in demonstrating a technique, rather
than finding the most efficient implementation.

\subsubsection{Constant diffusion}

For example, the simplest diagonal choice of \begin{equation}
{\cal D}_{\mathbf{j}\mathbf{j}'}=\beta J_{\mathbf{j}}\delta_{\mathbf{j}\mathbf{j}'}\end{equation}
 leads to the following stochastic equation for the link noise:

\begin{eqnarray}
\frac{\partial W_{ij}}{\partial\tau} & = & -\frac{1}{2}J_{ij}\beta\frac{\partial V}{\partial W_{ij}}+\xi_{ij}\left(\tau\right)\nonumber \\
 & = & -W_{ij}+\frac{1}{2}J_{ij}\beta\left\{ m_{i}+m_{j}\right\} +\xi_{ij}\left(\tau\right)\,\,,\end{eqnarray}
 where $m_{i}=\tanh\left(R_{i}\right)=\tanh\left(h_{i}\beta+\sum_{j}\left(W_{ij}\left(\beta\right)+W_{ji}\left(\beta\right)\right)\right)$,
and: \begin{equation}
\left\langle \xi_{ij}\left(\tau\right)\xi_{i'j'}\left(\tau'\right)\right\rangle =\beta J_{ij}\delta_{ii'}\delta_{jj'}\delta\left(\tau-\tau'\right).\end{equation}
 Changing variables to $R_{i}=h_{i}\beta+\sum_{j}W_{ij}^{+}\left(\beta\right)$,
with corresponding noises $\xi_{i}=\sum_{j}\left(\xi_{ij}+\xi_{ji}\right)$,
and an effective gain of $g_{i}=\beta\sum_{j}J_{ij}$, this reduces
to\begin{equation}
\frac{\partial R_{i}}{\partial\tau}=-R_{i}+g_{i}\tanh\left(R_{i}\right)+\beta h_{i}+\sum_{j}\beta J_{ij}\tanh\left(R_{j}\right)+\xi_{i}\left(\tau\right)\,\,.\label{eq:stochastic}\end{equation}

The important feature of this exact equation is that no additional
weighting is required. Each link noise equation is well localized,
only scaling with the total lattice size. That is, for a $D$-dimensional
lattice and nearest neighbour couplings, there are just $MD$ link
equations for $M$ lattice points. The algorithm can be improved further
by implementing link noises with variable correlation lengths for
calculations near the critical point, in order to spin-flip large
clusters more quickly, and to reduce the problem of critical slowing-down.
This could be achieved by having larger noise coefficients for longer
wavelength Fourier coefficients.

One can understand the equations physically as having a similar behaviour
to the equation for the gain of a laser, with the first term causing
loss and the second term gain, although with a nonlinear saturation
as well. The first term is dominant at high temperature (small $\beta$),
while the second term dominates at low temperature (large $\beta$).
The external magnetic field term is like an injected field in the
laser equations. The fourth describes correlations, while the last
is a noise term.

\subsection{Example:}

As an example, consider the uniform two-node case again, where there
is only one link and the two stochastic variables are perfectly correlated.
The stochastic equation is then: \begin{equation}
\frac{\partial R}{\partial\tau}=-R+2\beta J\tanh\left(R\right)+\beta h+\xi\left(\tau\right)\,\,,\label{eq:stochastcequation_twosite}\end{equation}
 with \begin{equation}
\left\langle \xi\left(\tau\right)\xi\left(\tau'\right)\right\rangle =2\beta J\delta\left(\tau-\tau'\right).\end{equation}

The correlation function is calculated from \begin{equation}
\left\langle \widehat{\sigma}_{1}^{z}\widehat{\sigma}_{2}^{z}\right\rangle =\left\langle \left[\tanh\left(R\right)\right]^{2}\right\rangle .\end{equation}

\begin{figure}
\includegraphics[width=8cm]{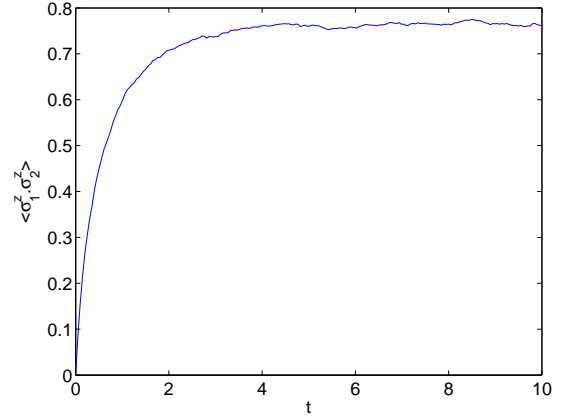}

\caption{\label{fig:Stochastic-calculation-of}Stochastic calculation of two
spin correlation: $J\beta=1$; $4000$ trajectories; step-size $.05$;
semi-implicit method with $3$ iterations.}

\end{figure}

\begin{figure}
\includegraphics[width=8cm]{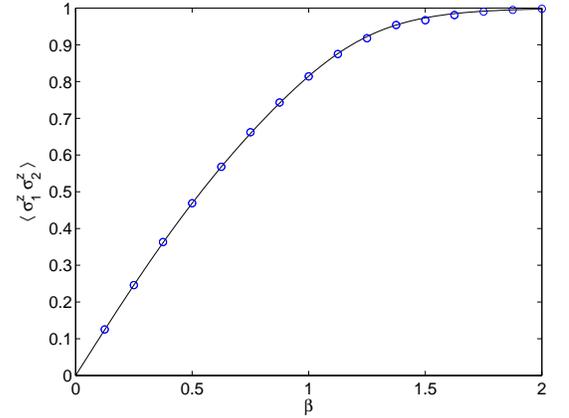}

\caption{\label{fig:Stochastic-calculation-over range}Stochastic calculation
of two spin correlations over a range of $J\beta$; comparison to
exact results.}

\end{figure}

The results of a simulation of Eq (\ref{eq:stochastcequation_twosite})
are shown in Fig (\ref{fig:Stochastic-calculation-of}). The corresponding
correct result for the two-spin correlation is given by Eqs (\ref{eq:exact_two-site})
and (\ref{eq:exact-two-site-stoch}) as: $\tanh(J\beta)=\tanh(1)=0.7616$.
Detailed results over a range of temperatures are compard with exact
results at thermal equilibrium in Fig (\ref{fig:Stochastic-calculation-over range}).

\begin{figure}
\includegraphics[width=8cm]{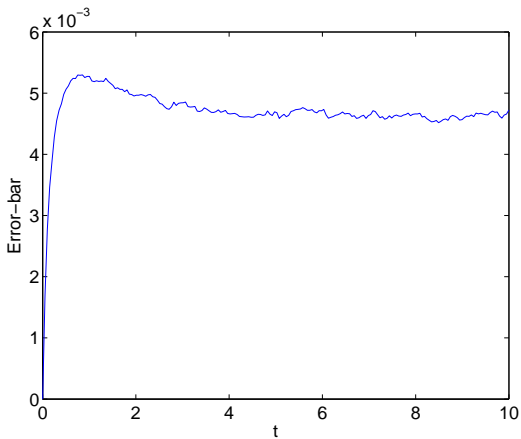}

\caption{\label{fig:Sampling-error-of}Sampling error of two spin correlations.}

\end{figure}

The sampling error in an ensemble of ${\cal N}$ trajectories can
be estimated as $\sigma/\sqrt{{\cal N}}$, where $\sigma$ is the
standard deviation of the calculated results, and assuming a nearly
normal distribution. The actual sampling error for this simulation
varies in time, and was estimated as $0.005$, for large times --
near equilibrium -- as shown in Fig (\ref{fig:Sampling-error-of}).

Given this estimated error, the calculated stochastic result for the
correlation agrees with the exact solution within the sampling error.

\subsection{Two-dimensional lattice calculation}

As a non-trivial example calculation, we consider a $10\times10$
Ising model with periodic boundary conditions. Couplings are nearest
neighbor, on a rectangular lattice with $J_{ij}=1$ and $h=0$.

\begin{figure}
\includegraphics[width=8cm]{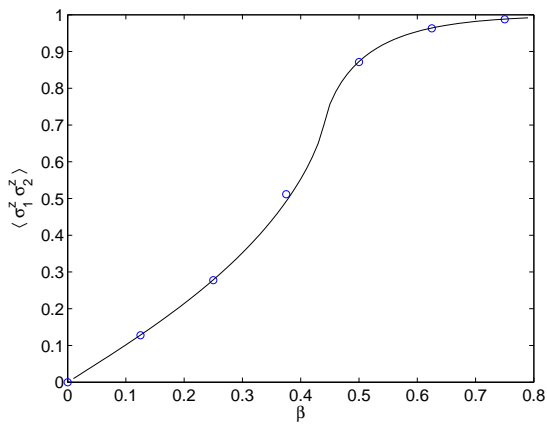}

\caption{Nearest-neighbor correlations for a $10\times10$ lattice as a function
of inverse temperature. Circles: Results from stochastic calculations,
1000 trajectories. Solid line: Exact solution in the limit of an infinite
lattice.}

\end{figure}

The numerically calculated nearest-neighbour correlation function
is given for six different inverse temperatures $\beta$. Once the
relevant stochastic averages have reached steady-state, they are time-averaged
as well as stochastically-averaged to give the correlation functions.

The results are shown in Figure (3), along with a comparison to the
known exact solution\cite{Onsager} in the limit of an infinite lattice.
The critical inverse temperature in this case is $\beta_{c}\approx0.44$,
as seen in the exact solution.

\section{Summary}

We have shown how to obtain a general phase-space representation with
positive-definite diffusion, for multiple SU(2) and more general SU($n$)
quantum systems, with couplings obtained from the corresponding operator
algebra. In the case of qubits or two-level systems, the appropriate
operator algebra is the spin half SU(2) algebra. This allows some
further simplifications in obtaining evolution equations.

The main application of these methods is to obtain stochastic methods
for calculating either canonical ensembles or time-evolution of coupled
atomic or spin systems. We have taken the exactly soluble Ising model
as an example. The resulting stochastic equations were solved for
correlation functions at finite temperature, and we found excellent
agreement with known exact results. These techniques can also be applied
to more complex n-level cases, with time-evolution and coupling to
external reservoirs.

\begin{acknowledgments}
Funding for this research was generously provided by the Australian
Research Council Center of Excellence program. 
\end{acknowledgments}


\begin{thebibliography}{10}
\bibitem{Schrod}E. Schrödinger, Naturwissenschaften \textbf{14},
664, 1926.

\bibitem{Dicke}R. H. Dicke, Phys. Rev. \textbf{93}, 99 (1954).

\bibitem{Radcliffe}J. M. Radcliffe, J. Phys. A \textbf{4}, 313 (1971).

\bibitem{ACGT}F. T. Arecchi, E. Courtens, R. Gilmore, and H. Thomas,
Phys. Rev. A \textbf{6}, 2211 (1972).

\bibitem{Perelemov}A. Perelomov, Commun. Math. Phys. \textbf{26,}
222 (1972); K. Nemoto, J. Phys. A: Math. Gen. \textbf{33}, 3493 (2000).

\bibitem{Narducci}L. M. Narducci, C. A. Coulter, and C. M. Bowden,
Phys. Rev. A \textbf{9}, 829 (1974).

\bibitem{Gilmore}R. Gilmore, C. M. Bowden, and L. M. Narducci, Phys.
Rev. A \textbf{12}, 1019 (1975).

\bibitem{Shastry}B. Sriram Shastry, G. S. Agarwal, and I. Rama Rao,
Pramana \textbf{11}, 85 (1978).

\bibitem{Lee}C. T. Lee, Phys. Rev. A \textbf{30}, 3308 (1984).

\bibitem{Drummond_SU(N)}P. D. Drummond, Phys. Letts. A \textbf{106},
118 (1984).

\bibitem{GitmanSU(n)Completeness}D. M. Gitman and A. L. Shelepin,
J. Phys A \textbf{26}, 313 (1993).

\bibitem{Wig-Wigner}E.~P.~Wigner, Phys.~Rev. \textbf{40}, 749
(1932).

\bibitem{Hus-Q}K.~Husimi, Proc.~Phys.~Math.~Soc.~Jpn. \textbf{22},
264 (1940).

\bibitem{GSp} R. J. Glauber, Phys. Rev.~\textbf{131}, 2766 (1963);
E. C. G. Sudarshan, Phys. Rev. Lett.~\textbf{10}, 277 (1963).

\bibitem{CG-Q}K.~E.~Cahill and R.~J.~Glauber, Phys.~Rev. \textbf{177},
1882 (1969).

\bibitem{Agarwal}G. S. Agarwal and E. Wolf, Phys. Rev. \textbf{D
2} , 2161 (1970).

\bibitem{positiveP1}S.~Chaturvedi, P.~D.~Drummond, and D.~F.~Walls,
J.~Phys.~A \textbf{10}, L187-192 (1977).

\bibitem{pp}P. D. Drummond, C. W. Gardiner, J.~Phys.~A:~Math.~Gen.~\textbf{13},
2353 (1980).

\bibitem{QuantNoise}C. W. Gardiner, \emph{Quantum Noise,} (Springer-Verlag,
Berlin, 1991).

\bibitem{GaugeP}P. Deuar and P. D. Drummond, Phys. Rev. A \textbf{66},
033812 (2002).

\bibitem{ChatWig}S. Chaturvedi, E. Ercolessi, G. Marmo, G. Morandi,
N. Mukunda amd R. Simon, J. Phys A. \textbf{39} 1405 (2006).

\bibitem{SmithGard}A. M. Smith and C. W. Gardiner, Phys. Rev. A.
\textbf{38}, 4073 (1988).

\bibitem{carusottothermo}Y.~Castin, and I.~Carusotto, J.~Phys.~B
\textbf{34}, 4589 (2001).

\bibitem{Pathint}E. A. Kochetov, Phys. Rev. B \textbf{52}, 4402 (1995).

\bibitem{white92}S. R. White, Phys. Rev. Lett. \textbf{69}, 2863
(1992).

\bibitem{vidal03}G. Vidal, Phys. Rev. Lett. \textbf{91}, 147902 (2003).

\bibitem{verstraete04}F. Verstraete, D. Porras and J.I. Cirac, Phys.
Rev. Lett. 93, 227205 (2004).

\bibitem{Ising} E. Ising, Z. Phys. \textbf{31}, 253 (1925).

\bibitem{Landau}L.D. Landau, E.M. Lifshitz, \emph{Statistical physics}
(Pergamon Press, 1958).

\bibitem{Onsager}L. Onsager, Phys. Rev. \textbf{65,} 117 (1944).

\bibitem{Baxter}R.J. Baxter, \emph{Exactly solved models in statistical
mechanics} , (Acad. Press, 1982), F.Y. Wu, Rev. Mod. Phys. \textbf{54},
235 (1982).

\bibitem{Fisher}M. E. Fisher, Rev. Mod. Phys. \textbf{46}, 597 (1974).

\bibitem{BinderYoung}K. Binder and A.P. Young, Rev. Mod. Phys., \textbf{58,}
801 (1986).

\bibitem{Percolation}D. Stauffer, A. Aharony, \emph{Introduction
to percolation theory} , (Taylor \& Francis, 1992).

\bibitem{Jaksch}D. Jaksch, C. Bruder, J. I. Cirac, C. W. Gardiner
and P. Zoller, Phys. Rev. Lett. \textbf{81}, 3108 (1998).

\bibitem{wilson:74}K.~G.~Wilson, Phys.~Rev.~D \textbf{10}, 2445
(1974); D.~M.~Ceperley, Rev.~Mod.~Phys. \textbf{67}, 279 (1995).

\bibitem{Binder}K. Binder ed, \emph{Monte Carlo methods in statistical
physics} , (Springer, Berlin, 1979).

\bibitem{Swendsen}R.H. Swendsen and J. S. Wang, Phys. Rev. Lett.
\textbf{58}, 86 (1987).

\bibitem{BloteHeringa}H.W.J. Blote, J.R. Heringa and E. Luijten,
Computer Physics Communications \textbf{147}, 58 (2002); H.W.J. Blote,
J.R. Heringa and M.M. Tsypin, Physical Review E \textbf{62}, 77 (2000);
J.R. Heringa and H.W.J. Blote, Physical Review E \textbf{57}, 4976
(1998).

\bibitem{Metropolis}N. Metropolis, A. Rosenbluth, M. Rosenbluth,
A. Teller and E. Teller, J. Chem. Phys. \textbf{21}, 1087 (1953).

\bibitem{Hamermesh}M. Hamermesh, \emph{Group Theory} (Addison-Wesley,
1962).

\bibitem{DDK2004}P.D. Drummond, P. Deuar, and K.V. Kheruntsyan, Phys.
Rev. Lett. \textbf{92}, 040405 (2004).

\bibitem{GGD-Validity}A.~Gilchrist, C.~W.~Gardiner, and P.~D.~Drummond,
Phys.~Rev.~A \textbf{55}, 3014 (1997).

\bibitem{Collecut}G. R. Collecutt and P. D. Drummond, Comput. Phys.
Commun. \textbf{142}, 219-223 (2001).

\bibitem{xmds}For example, see website www.xmds.org. At the time
of writing, this makes available an automatic code generator known
as XMDS using clustering (parallel) technology, available under a
public license. 
\end{thebibliography}
\end{document}